\documentclass[aps,showpacs,preprint]{revtex4}
\usepackage{graphicx}
\begin{document}
\title{\bf Simulation of a Heisenberg XY- chain and realization of a perfect state transfer algorithm using
liquid nuclear magnetic resonance \footnote{Corresponding authors:
Jingfu Zhang, zhangjfu2000@yahoo.com  and Gui Lu Long,
gllong@mail.tsinghua.edu.cn}}
\author{\small{} Jingfu Zhang$^{1}$, Gui Lu
Long$^{1,4}$, Wei Zhang$^{2}$, Zhiwei Deng$^{2}$, Wenzhang
Liu$^{1}$, and Zhiheng Lu$^{3}$ }
\address{$^{1}$Key Laboratory For Quantum Information and Measurements of MOE,  and
Department of Physics, Tsinghua University, Beijing,
100084, P R China\\
 $^{2}$Testing and Analytical Center, Beijing
Normal University,
 Beijing, 100875, P R China\\
$^{3}$Department of Physics, \small{}Beijing Normal University,
Beijing, 100875, P R China \\
$^{4}$ Key Laboratory For Atomic and Molecular Nanosciences,
Tsinghua University, Beijing 100084, P R China }
\date{\today}
\begin{abstract}
   The three- spin chain with Heisenberg XY- interaction is simulated in
a three- qubit nuclear magnetic resonance (NMR) quantum computer.
The evolution caused by the XY- interaction is decomposed into a
series of single- spin rotations and the $J$- coupling evolutions
between the neighboring spins. The perfect state transfer (PST)
algorithm proposed by M. Christandl et al [Phys. Rev. Lett, 92,
187902(2004)] is realized in the XY- chain.
\end{abstract}

\maketitle
\section{Introduction}
   In 1982, R. P. Feynman proposed the idea of quantum computer,
and pointed out that a quantum computer can simulate physics more
efficiently than its classical counterpart \cite{Feynman82}. The
information carriers for quantum computation are qubits. Unlike a
classical bit, a qubit can lie in a superposition of two states,
according to the quantum mechanical principle of superposition.
Essentially, the superpositions of quantum states lead to the
advantages of the quantum computers over the classical computers.
The quantum gates and quantum networks proposed by D. Deutsch
provide a convenient method for people to think about how to build
a quantum computer in a similar way to build a classical computer
\cite{Deutsch85,Deutsch89}. His work can be thought as a milestone
in the history of quantum computer\cite{Steane98}. The current
quantum network theory has shown that it is possible to construct
an arbitrary $n$-qubit quantum gate by using only a finite set of
single- qubit gates and two- qubit gates \cite{4,5}, so that these
basic quantum gates are universal for quantum computation
\cite{6,7,8}. The prime factorization algorithm proposed by P. W.
Shor \cite{Shor94} and the quantum search algorithm proposed by L.
K. Grover \cite{Grover97} show the potential advantages of the
quantum computers and accelerate the development of quantum
computation.

  There are several physical systems that can implement quantum computation.
They are liquid nuclear magnetic resonance (NMR)
\cite{Vandersypen04}, quantum dots \cite{Loss98,Loss99}, solid NMR
\cite{Kane,Ladd,Suter}, electron spins \cite{Vrijen}, trapped ions
\cite{Zoller95,Zoller00}, superconduction qubits
\cite{Makhlin,Devoret}, and cavity QED systems
\cite{Sleator,Gigovannetti,Yamaguchi,Zheng}. Because of its
technologic maturation and convenience in manipulation, liquid NMR
has been an important experimental method to implement quantum
algorithm, error-correcting code, and simulate quantum systems
\cite{2,Boulant,Miquel,Lloyd,Boghosian,Somaroo,Du1,Du2,zhangpra,zhangjopb,Peng,Negrevergne,Yang}.

  In the above systems, the interactions between qubits, at least
between the neighboring qubits, are necessary for quantum
computation. The Heisenberg interaction naturally exists in the
various spin systems. In the liquid NMR system, the Heisenberg
interaction exists in form of Ising interaction \cite{Gunlycke}.
In the other systems, the Heisenberg interaction takes more
various forms. D. P. DiVincenzo et al pointed out that the
Heisenberg interaction alone can be universal for quantum
computation, if the coded qubit states are introduce
\cite{DiVincenzo,Bacon,Knill}. This result is exciting, because
the single- spin operations, which usually cause additional
difficulties in manipulations in some systems, can be avoided. The
perfect state transfer (PST) algorithm proposed by M. Christandl
et al satisfies such a condition that no single-spin operations
are needed \cite{Christandl}. The algorithm can transfer an
arbitrary quantum state between the two ends of a spin chain or a
more complex spin network in a fixed period time only using the
XY- interaction. If the state is transferred in a more than three
spin chain, the coded qubits are needed, so that the chain is
extended to a network. Compared with the state transfer based on
SWAP operations, where single-spin operations are used to switch
on or off the couplings between spins \cite{Madi}, the PST
algorithm is easy to implement in some solid systems.

  The Heisenberg interaction is expected to play an important role
in building large- scale quantum computers, and it has become an
interesting topic in the field of quantum information. M. C.
Arnesen et al' work indicated that the quantum entanglement phase
transition occurs in the one dimension Heisenberg model
\cite{Arnesen}. L. Zhou et al pointed out that the thermal
entanglement can be enhanced in an anisotropic Heisenberg XYZ
chain\cite{Zhou}. J. P. Keating et al separated a quantum spin
chain into two parts, and computed the entropy of entanglement
between them \cite{Keating}. M. Mohseni et al proposed a
fault-tolerant quantum computation using Heisenberg interactions
\cite{Mohseni}. The other issues related to the Heisenberg model,
such as the Heisenberg chain with the next-nearest-neighbor
interaction, are also discussed \cite{Gu04,Gu03,Wang04,Wang02}. In
experiment, the quantum entanglement phase transition in a two-
 spin Ising- chain was demonstrated in an NMR quantum
computer \cite{Peng}.

  Quantum simulation has been an interesting topic since the quantum computer
is born\cite{Feynman82,Lloyd,Boghosian}. Liquid NMR has displayed
its powerful ability to simulate quantum systems, and various of
quantum systems have been successfully simulated in NMR quantum
computers \cite{Somaroo,Peng,Negrevergne,Yang}. In this paper, we
simulate a three- spin XY- chain with the Heisenberg interaction
and realize the perfect state transfer(PST) algorithm
\cite{Christandl} using a liquid NMR quantum computer.

\section{Simulating the three- spin XY- chain using liquid NMR}

  The Hamiltonian for a three spin XY- chain with the neighboring Heisenberg interaction is

 \begin{equation}\label{xy}
    H_{XY}=\frac{1}{2} J (\sigma_{x}^{1}\sigma_{x}^{2}+\sigma_{y}^{1}\sigma_{y}^{2}
    +\sigma_{x}^{2}\sigma_{x}^{3}+\sigma_{y}^{2}\sigma_{y}^{3}),
\end{equation}
where $\sigma_{x/y}^{j}(j=1,2,3)$ are the Pauli matrices for the
angular momentum of the spins, and $J$ is the coupling constant
between two spins. For convenience in expression, $\hbar$ has been
set to $1$. The evolution caused by $H_{XY}$ can be expressed as

\begin{equation}\label{Uxy}
   U(t)=e^{-iH_{XY} t},
\end{equation}
where $t$ is the evolution time. In order to represent $U(t)$ as a
liquid NMR version, we introduce two commutable operators
$A=(\sigma_{x}^{1}\sigma_{x}^{2}+\sigma_{y}^{2}\sigma_{y}^{3})/2$,
and
$B=(\sigma_{y}^{1}\sigma_{y}^{2}+\sigma_{x}^{2}\sigma_{x}^{3})/2$.
$U(t)$ can be rewrite as  $U(t)=U_{A}(t)U_{B}(t)$, where

\begin{equation}\label{UA}
   U_{A}(t)= e^{-iJtA}\equiv
   e^{-iJt(\sigma_{x}^{1}\sigma_{x}^{2}+\sigma_{y}^{2}\sigma_{y}^{3})/2},
\end{equation}

\begin{equation}\label{UB}
   U_{B}(t)= e^{-iJtB}\equiv
   e^{-iJt(\sigma_{y}^{1}\sigma_{y}^{2}+\sigma_{x}^{2}\sigma_{x}^{3})/2}.
\end{equation}

  We define three operators $L_{x}^{A}\equiv \sigma_{x}^{1}\sigma_{x}^{2}/2$,
  $L_{y}^{A}\equiv \sigma_{y}^{2}\sigma_{y}^{3}/2$, and
$L_{z}^{A}\equiv \sigma_{x}^{1}\sigma_{z}^{2}\sigma_{y}^{3}/2$.
These three operators can be viewed as the three components of the
angular momentum vector denoted by  $\bf{L^{A}}$, because they
satisfy the commuting conditions
$[L_{x}^{A},L_{y}^{A}]=iL_{z}^{A}$,
$[L_{y}^{A},L_{z}^{A}]=iL_{x}^{A}$, and
$[L_{z}^{A},L_{x}^{A}]=iL_{y}^{A}$. Eq. (\ref{UA}) can be
rewritten as

\begin{equation}\label{UAr}
    U_{A}(t)=e^{-iJt (L_{x}^{A}+L_{x}^{B})}=e^{-i(\sqrt{2}Jt)
\bf{L^{A}}\cdot \bf{n}},
\end{equation}
where the vector ${\bf n}=(1/\sqrt{2},1/\sqrt{2},0)$, and it
denotes the direction of the rotation axis for $U_{A}(t)$. The
separate angles between $\bf{n}$ and $x$, $y$, $z$ axes are
$\pi/4$, $\pi/4$, and $\pi/2$, respectively. Using the theories of
angular momentum, we obtain

\begin{eqnarray}\label{UAf}
 U_{A}(t)&=&e^{-i \frac{\pi}{4}L_{z}^{A}}e^{-i\sqrt{2}Jt L_{x}^{A}}e^{i \frac{\pi}{4}
L_{z}^{A}}\nonumber\\
 &=& e^{-i\frac{\pi}{8}\sigma_{x}^{1}\sigma_{z}^{2}\sigma_{y}^{3}}
 e^{-i(\frac{Jt}{\sqrt{2}})\sigma_{x}^{1}\sigma_{x}^{2}}
e^{i\frac{\pi}{8}\sigma_{x}^{1}\sigma_{z}^{2}\sigma_{y}^{3}}\nonumber\\
&=&\cos(\frac{Jt}{\sqrt{2}})-\frac{i}{\sqrt{2}}\sin(\frac{Jt}{\sqrt{2}})(\sigma_{x}^{1}\sigma_{x}^{2}
 +\sigma_{y}^{2}\sigma_{y}^{3}).
\end{eqnarray}

  In a similar way, through defining
  $L_{x}^{B}\equiv
\sigma_{x}^{2}\sigma_{x}^{3}/2$,
  $L_{y}^{B}\equiv \sigma_{y}^{1}\sigma_{y}^{2}/2$, and
$L_{z}^{B}\equiv \sigma_{y}^{1}\sigma_{z}^{2}\sigma_{x}^{3}/2$ as
the three components of the angular momentum vector denoted as
$\bf{L^{B}}$, we obtain

\begin{eqnarray}\label{UBf}
 U_{B}(t)&=&e^{-i \frac{\pi}{4}L_{z}^{B}}e^{-i\sqrt{2}Jt L_{x}^{B}}e^{i \frac{\pi}{4}
L_{z}^{B}}\nonumber\\
 &=&e^{-i\frac{\pi}{8}\sigma_{y}^{1}\sigma_{z}^{2}\sigma_{x}^{3}}
 e^{-i(\frac{Jt}{\sqrt{2}})\sigma_{x}^{2}\sigma_{x}^{3}}
e^{i\frac{\pi}{8}\sigma_{y}^{1}\sigma_{z}^{2}\sigma_{x}^{3}}\nonumber\\
&=&\cos(\frac{Jt}{\sqrt{2}})-\frac{i}{\sqrt{2}}\sin(\frac{Jt}{\sqrt{2}})(\sigma_{y}^{1}\sigma_{y}^{2}
 +\sigma_{x}^{2}\sigma_{x}^{3}).
\end{eqnarray}
One can prove the last equations in Eqs.(\ref{UAf}) and
(\ref{UBf}) directly through Eqs. (\ref{UA}) and (\ref{UB}) using
$((\sigma_{x}^{1}\sigma_{x}^{2}
 +\sigma_{y}^{2}\sigma_{y}^{3})/\sqrt{2})^2=1$, and
$((\sigma_{y}^{1}\sigma_{y}^{2}
 +\sigma_{x}^{2}\sigma_{x}^{3})/\sqrt{2})^2=1$.
Combining Eqs. (\ref{UAf}) and (\ref{UBf}), we obtain

\begin{equation}\label{Uf}
  U(t)=e^{-i\frac{\pi}{8}\sigma_{x}^{1}\sigma_{z}^{2}\sigma_{y}^{3}}
 e^{-i(\frac{Jt}{\sqrt{2}})\sigma_{x}^{1}\sigma_{x}^{2}}
e^{i\frac{\pi}{8}\sigma_{x}^{1}\sigma_{z}^{2}\sigma_{y}^{3}}
e^{-i\frac{\pi}{8}\sigma_{y}^{1}\sigma_{z}^{2}\sigma_{x}^{3}}
 e^{-i(\frac{Jt}{\sqrt{2}})\sigma_{x}^{2}\sigma_{x}^{3}}
e^{i\frac{\pi}{8}\sigma_{y}^{1}\sigma_{z}^{2}\sigma_{x}^{3}}.
\end{equation}
Each of the six factors in Eq. (\ref{Uf}) can be realized using
liquid NMR. Consequently the three- spin XY- chain can be
simulated in a three- spin liquid NMR system.

\section{Implementing the perfect state transfer algorithm in the XY- chain}
   The PST algorithm was proposed by M. Christandl et al \cite{Christandl},
and it can be implemented in the XY chain. The algorithm can
transfer an arbitrary quantum state between the two ends of the
chain in a fixed period time, only using the XY- interaction.
Unlike the state transfer based on SWAP operations \cite{Madi},
the PST algorithm do not require single- spin operations. Hence
the algorithm is more feasible to realize in some systems, such as
the electron-spin-resonance system, where the single- spin
operations cause many experimental difficulties \cite{Vrijen}.

Letting $\varphi\equiv Jt/\sqrt{2}$, Eq. (\ref{Uf}) is represented
as the matrix

\begin{equation}\label{Ufm}
U(t)=\left (\begin{array}{cccccccc}
  1 & 0 & 0 & 0 & 0 & 0 & 0 & 0 \\
  0 & \cos^{2}\varphi & -\frac{i}{\sqrt{2}}\sin(2\varphi) & 0 & -\sin^{2}\varphi & 0 & 0 & 0 \\
  0  & -\frac{i}{\sqrt{2}}\sin(2\varphi)  & \cos(2\varphi)  &  0 & -\frac{i}{\sqrt{2}}\sin(2\varphi)  & 0  &  0 & 0  \\
  0  &  0 &  0 & \cos^{2}\varphi  &0   &   -\frac{i}{\sqrt{2}}\sin(2\varphi)&  -\sin^{2}\varphi &0   \\
  0  & -\sin^{2}\varphi  &  -\frac{i}{\sqrt{2}}\sin(2\varphi) & 0  & \cos^{2}\varphi  &  0 & 0  & 0  \\
  0  & 0  & 0  &-\frac{i}{\sqrt{2}}\sin(2\varphi)   &  0 & \cos(2\varphi)  &  -\frac{i}{\sqrt{2}}\sin(2\varphi)  &  0 \\
  0  & 0  & 0  &  -\sin^{2}\varphi & 0  &  -\frac{i}{\sqrt{2}}\sin(2\varphi) & \cos^{2}\varphi  & 0  \\
  0  & 0  &  0 & 0  & 0  &  0 & 0  & 1  \\
\end{array}\right ).
\end{equation}
The order of the basis states is  $|000\rangle$, $|001\rangle$,
$|010\rangle$,  $|011\rangle$,  $|100\rangle$, $|101\rangle$,
$|110\rangle$,  $|111\rangle$, where $|0\rangle$ and $|1\rangle$
denote the spin up and down states, respectively. When
$t=\frac{\pi}{\sqrt{2}J}$, one obtains

\begin{equation}\label{PST}
U(\frac{\pi}{\sqrt{2}J})=\left (\begin{array}{cccccccc}
  1 & 0 & 0 & 0 & 0 & 0 & 0 & 0 \\
  0 & 0 & 0 & 0 & -1 & 0 & 0 & 0 \\
  0  & 0  & -1  &  0 & 0  & 0  &  0 & 0  \\
  0  &  0 &  0 & 0  &0   &   0&  -1 &0   \\
  0  & -1  &  0 & 0  & 0  &  0 & 0  & 0  \\
  0  & 0  & 0  &0   &  0 & -1  &  0  &  0 \\
  0  & 0  & 0  &  -1 & 0  &  0 & 0  & 0  \\
  0  & 0  &  0 & 0  & 0  &  0 & 0  & 1  \\
\end{array}\right ).
\end{equation}
Obviously, $U|000\rangle=|000\rangle$,
$U|001\rangle=-|100\rangle$, $U|010\rangle=-|010\rangle$,
$U|011\rangle=-|110\rangle$, $U|100\rangle=-|001\rangle$,
$U|101\rangle=-|101\rangle$, $U|110\rangle=-|011\rangle$, and
$U|111\rangle=|111\rangle$. We use
$|\psi\rangle_{in}=(\alpha|0\rangle+\beta|1\rangle)|00\rangle $ as
the input state by setting spin $1$ into state
$(\alpha|0\rangle+\beta|1\rangle)$, where $\alpha$, $\beta$ are
two arbitrary complex numbers. $U(\frac{\pi}{\sqrt{2}J})$
transforms $|\psi\rangle_{in}$ to
$|00\rangle(\alpha|0\rangle-\beta|1\rangle)$, where spin $3$ lies
in state $(\alpha|0\rangle-\beta|1\rangle)$, and the perfect state
transfer is completed. Obviously, a simple operation $\sigma_{z}$
can transform $(\alpha|0\rangle-\beta|1\rangle)$ to
$(\alpha|0\rangle+\beta|1\rangle)$, and therefore
$|\psi\rangle_{in}$ is transferred from spin $1$ to spin $3$.

  The implementation of PST algorithm in two- or three- spin chain
does not require the coded qubits. However in more than three-spin
chain, the coded qubits are needed to design so as to extend the
chain to a more complex network. The details can been found in
\cite{Christandl}.

\section{Realization in a three- NMR quantum computer}
  The experiment uses a sample of Carbon-13 labelled
trichloroethylene (TCE) dissolved in d-chloroform. Data are taken
with a Bruker DRX 500 MHz spectrometer. The temperature is
controlled at 22$^{\circ}C$. $^{1}$H is denoted as qubit 2, the
$^{13}$C directly connecting to $^{1}$H is denoted as qubit 1, and
the other $^{13}$C is denoted as qubit 3. The three qubits are
denoted as C1, H2 and C3. The Hamiltonian of the three-qubit
system is \cite{s10}

\begin{equation}\label{nmr}
  H_{NMR}=-\pi\nu_{1}\sigma_{z}^{1}-\pi\nu_{2}\sigma_{z}^{2}-\pi\nu_{3}\sigma_{z}^{3}
  +\frac{1}{2}\pi J_{12}\sigma_{z}^{1}\sigma_{z}^{2}+\frac{1}{2}\pi J_{23 }\sigma_{z}^{2}\sigma_{z}^{3}
  +\frac{1}{2}\pi J_{13} \sigma_{z}^{1}\sigma_{z}^{3},
\end{equation}
where  $\nu_{1}$, $\nu_{2}$, $\nu_{3}$ are the resonance
frequencies of C1, H2 and C3, and $\nu_{3}=\nu_{1}+904.4$Hz. The
coupling constants are measured to be $J_{12}=200.9$ Hz,
$J_{23}=9.16$ Hz, and $J_{13}=103.1$Hz. The coupled-spin evolution
between two spins is denoted as
\begin{equation}\label{2}
  [\tau_{jl}]=e^{-i\frac{1}{2}\pi J_{jl} \tau \sigma_{z}^{j} \sigma_{z}^{l}},
\end{equation}
where $l=1,2,3$, and $j\neq l$. $[\tau_{jl}]$ can be realized by
averaging the coupling constants other than $J_{jl}$ to
zero\cite{s15}.

  The three- body and two- body interactions in Eq. (\ref{Uf}) can be
expressed as \cite{DUPRA03}

\begin{equation}\label{xzy}
   e^{-i\frac{\pi}{8}\sigma_{x}^{1}\sigma_{z}^{2}\sigma_{y}^{3}}
   =e^{-i\frac{\pi}{4}\sigma_{y}^{1}}e^{i\frac{\pi}{4}\sigma_{x}^{3}}
   e^{-i\frac{\pi}{8}\sigma_{z}^{1}\sigma_{z}^{2}\sigma_{z}^{3}}
   e^{i\frac{\pi}{4}\sigma_{y}^{1}}e^{-i\frac{\pi}{4}\sigma_{x}^{3}},
\end{equation}

\begin{equation}\label{xzyp}
   e^{i\frac{\pi}{8}\sigma_{x}^{1}\sigma_{z}^{2}\sigma_{y}^{3}}
   =e^{-i\frac{\pi}{4}\sigma_{y}^{1}}e^{i\frac{\pi}{4}\sigma_{x}^{3}}
   e^{i\frac{\pi}{8}\sigma_{z}^{1}\sigma_{z}^{2}\sigma_{z}^{3}}
   e^{i\frac{\pi}{4}\sigma_{y}^{1}}e^{-i\frac{\pi}{4}\sigma_{x}^{3}},
\end{equation}

\begin{equation}\label{yzx}
   e^{-i\frac{\pi}{8}\sigma_{y}^{1}\sigma_{z}^{2}\sigma_{x}^{3}}
   =e^{i\frac{\pi}{4}\sigma_{x}^{1}}e^{-i\frac{\pi}{4}\sigma_{y}^{3}}
   e^{-i\frac{\pi}{8}\sigma_{z}^{1}\sigma_{z}^{2}\sigma_{z}^{3}}
   e^{-i\frac{\pi}{4}\sigma_{x}^{1}}e^{i\frac{\pi}{4}\sigma_{y}^{3}},
\end{equation}

\begin{equation}\label{yzxp}
   e^{i\frac{\pi}{8}\sigma_{y}^{1}\sigma_{z}^{2}\sigma_{x}^{3}}
   =e^{i\frac{\pi}{4}\sigma_{x}^{1}}e^{-i\frac{\pi}{4}\sigma_{y}^{3}}
   e^{i\frac{\pi}{8}\sigma_{z}^{1}\sigma_{z}^{2}\sigma_{z}^{3}}
   e^{-i\frac{\pi}{4}\sigma_{x}^{1}}e^{i\frac{\pi}{4}\sigma_{y}^{3}},
\end{equation}

\begin{equation}\label{x12}
e^{-i\varphi\sigma_{x}^{1}\sigma_{x}^{2}}=e^{-i\frac{\pi}{4}\sigma_{y}^{1}}e^{-i\frac{\pi}{4}\sigma_{y}^{2}}
e^{-i\varphi\sigma_{z}^{1}\sigma_{z}^{2}}
e^{i\frac{\pi}{4}\sigma_{y}^{1}}e^{i\frac{\pi}{4}\sigma_{y}^{2}},
\end{equation}

\begin{equation}\label{x23}
e^{-i\varphi\sigma_{x}^{2}\sigma_{x}^{3}}=e^{-i\frac{\pi}{4}\sigma_{y}^{2}}e^{-i\frac{\pi}{4}\sigma_{y}^{3}}
e^{-i\varphi\sigma_{z}^{2}\sigma_{z}^{3}}
e^{i\frac{\pi}{4}\sigma_{y}^{2}}e^{i\frac{\pi}{4}\sigma_{y}^{3}}.
\end{equation}
Through substituting Eqs. (\ref{xzy}-\ref{x23}) into Eq.
(\ref{Uf}), and after simplification, one obtains

\begin{eqnarray}\label{Ufs}
U(t)&=&e^{-i\frac{\pi}{4}\sigma_{y}^{1}}e^{i\frac{\pi}{4}\sigma_{x}^{3}}
   e^{-i\frac{\pi}{8}\sigma_{z}^{1}\sigma_{z}^{2}\sigma_{z}^{3}}
e^{-i\frac{\pi}{4}\sigma_{y}^{2}}e^{-i\varphi\sigma_{z}^{1}\sigma_{z}^{2}}
e^{i\frac{\pi}{4}\sigma_{y}^{2}}e^{i\frac{\pi}{8}\sigma_{z}^{1}\sigma_{z}^{2}\sigma_{z}^{3}}
e^{i\frac{\pi}{4}(\sigma_{x}^{1}+\sigma_{x}^{3})}e^{-i\frac{\pi}{4}\sigma_{y}^{3}}\nonumber\\
&& \times
e^{i\frac{\pi}{2}(\sigma_{z}^{1}+\sigma_{z}^{3})}e^{-i\frac{\pi}{4}\sigma_{z}^{1}}
e^{-i\frac{\pi}{8}\sigma_{z}^{1}\sigma_{z}^{2}\sigma_{z}^{3}}e^{-i\frac{\pi}{4}\sigma_{y}^{2}}
e^{-i\varphi\sigma_{z}^{2}\sigma_{z}^{3}}e^{i\frac{\pi}{4}\sigma_{y}^{2}}
e^{i\frac{\pi}{8}\sigma_{z}^{1}\sigma_{z}^{2}\sigma_{z}^{3}}
e^{-i\frac{\pi}{4}\sigma_{x}^{1}}e^{i\frac{\pi}{4}\sigma_{y}^{3}}.
\end{eqnarray}
In Eq. (\ref{Ufs}), $e^{i\frac{\pi}{4}\sigma_{y}^{2}}$ is realized
by a $\pi/2$ radio frequency (rf) pulse exciting H2 along
$y$-axis. Such a pulse is denoted by $[\pi/2]_{y}^{2}$. The
operation $e^{i\frac{\pi}{4}(\sigma_{x}^{1}+\sigma_{x}^{3})}$ is
realized by a nonselective pulse $[\pi/2]_{x}^{1,3}$, exciting C1
and C3 simultaneously. The widths of $[\pi/2]_{y}^{2}$ and
$[\pi/2]_{x}^{1,3}$ are so short that they can be ignored.
$e^{i\frac{\pi}{2}(\sigma_{z}^{1}+\sigma_{z}^{3})}$ is realized by
a pulse sequence

\begin{eqnarray}\label{z180}
[\pi]_{x}^{1,3}-[\pi]_{y}^{1,3},
\end{eqnarray}
where the time order is from left to right. The operation
selective for C1 or C3 can be realized by the established pulse
sequence \cite{Linden,s15,Geen,zhangpra}. For example,
$e^{i\frac{\pi}{4}\sigma_{x}^{1}}$ is realized by

\begin{eqnarray}\label{rx}
[\frac{\pi}{2}]_{x}^{1}=[\frac{\pi}{2}]_{y}^{1,3}-e^{i\frac{\pi}{4}\sigma_{z}^{1}}-[-\frac{\pi}{2}]_{y}^{1,3}.
\end{eqnarray}
According to C. H. Tseng et al's work \cite{Tseng},
$e^{-i\frac{\pi}{8}\sigma_{z}^{1}\sigma_{z}^{2}\sigma_{z}^{3}}$ is
realized by

\begin{eqnarray}\label{zzz}
[-\frac{\pi}{2}]_{x}^{2}-[-\pi]_{y}^{2}-[\frac{9}{2J_{12}}]-[\frac{\pi}{2}]_{y}^{2}
-[\frac{1}{4J_{23}}]-[\frac{\pi}{2}]_{y}^{2}-[\frac{9}{2J_{12}}]-[\frac{\pi}{2}]_{x}^{2},
\end{eqnarray}
and $e^{i\frac{\pi}{8}\sigma_{z}^{1}\sigma_{z}^{2}\sigma_{z}^{3}}$
is realized by

\begin{eqnarray}\label{zzzp}
[-\frac{\pi}{2}]_{x}^{2}-[-\pi]_{y}^{2}-[\frac{7}{2J_{12}}]-[\frac{\pi}{2}]_{y}^{2}
-[\frac{1}{4J_{23}}]-[\frac{\pi}{2}]_{y}^{2}-[\frac{7}{2J_{12}}]-[\frac{\pi}{2}]_{x}^{2}.
\end{eqnarray}
The case of only one proton in the sample makes the three- body
interactions realize much easy. One should note that the direct
coupling between C1 and C3 is not used.

  We choose the state

\begin{equation}\label{ini}
    \rho_{ini A}=\sigma_{y}^{1}
\end{equation}
as the initial state to simulate the XY- chain in the three-spin
system. The pulse sequence

\begin{eqnarray}\label{inip}
[\frac{\pi}{2}]_{y}^{2}-[\frac{\pi}{2}]_{y}^{3}-[grad]_{z}-[\frac{\pi}{2}]_{x}^{1}
\end{eqnarray}
transforms the system from the equilibrium

\begin{equation}\label{equ}
  \rho_{eq}=\gamma_{C}(I_{z}^{1}+ I_{z}^{3})+\gamma_{H}I_{z}^{2},
\end{equation}
to $\rho_{ini A}$ \cite{Tseng}, where $\gamma_{C}$ and
$\gamma_{H}$ denote the gyromagnetic ratios of $^{13}C$ and
$^{1}$H, and $[grad]_{z}$ denotes a gradient pulse along $z$-
axis. The irrelative overall factors have been ignored. Using
$[U_{B},\rho_{iniA}]=0$, we obtain $\rho_{A}(t)= U(t)\rho_{ini
A}U^{\dag}(t)=U_{A}(t)\rho_{iniA}U^{\dag}_{A}(t)$. In experiments,
we replace $U(t)$ by $U_{A}(t)$, in order to simplify experimental
procedure, and obtain

\begin{equation}\label{pt}
 \rho_{A}(t)= \sigma_{y}^{1}\cos^{2}\varphi+
 \sigma_{z}^{1}\sigma_{x}^{2}\frac{1}{\sqrt{2}}\sin(2\varphi)-
 \sigma_{z}^{1}\sigma_{z}^{2}\sigma_{y}^{3}\sin^{2}\varphi.
\end{equation}
When $t=\frac{\pi}{\sqrt{2}J}$,  one obtains
$\rho_{A}(\frac{\pi}{\sqrt{2}J})=-
 \sigma_{z}^{1}\sigma_{z}^{2}\sigma_{y}^{3}$, which means that the state
$\sigma_{y}$ has been transferred from C1 to C3. Similarly, if the
initial state is chosen as

\begin{equation}\label{iniB}
    \rho_{ini B}=\sigma_{x}^{1},
\end{equation}
we obtain

\begin{eqnarray}\label{ptB}
 \rho_{B}(t)&=& U(t)\rho_{ini
B}U^{\dag}(t)=U_{B}(t)\rho_{iniB}U^{\dag}_{B}(t)\nonumber\\
&=&\sigma_{x}^{1}\cos^{2}\varphi+
 \sigma_{z}^{1}\sigma_{y}^{2}\frac{1}{\sqrt{2}}\sin(2\varphi)-
 \sigma_{z}^{1}\sigma_{z}^{2}\sigma_{x}^{3}\sin^{2}\varphi.
\end{eqnarray}
Obviously, $\rho_{B}(\frac{\pi}{\sqrt{2}J})=-
 \sigma_{z}^{1}\sigma_{z}^{2}\sigma_{x}^{3}$, which means that $\sigma_{x}$ has been
transferred from C1 to C3.

\begin{figure}
\includegraphics[width=4in]{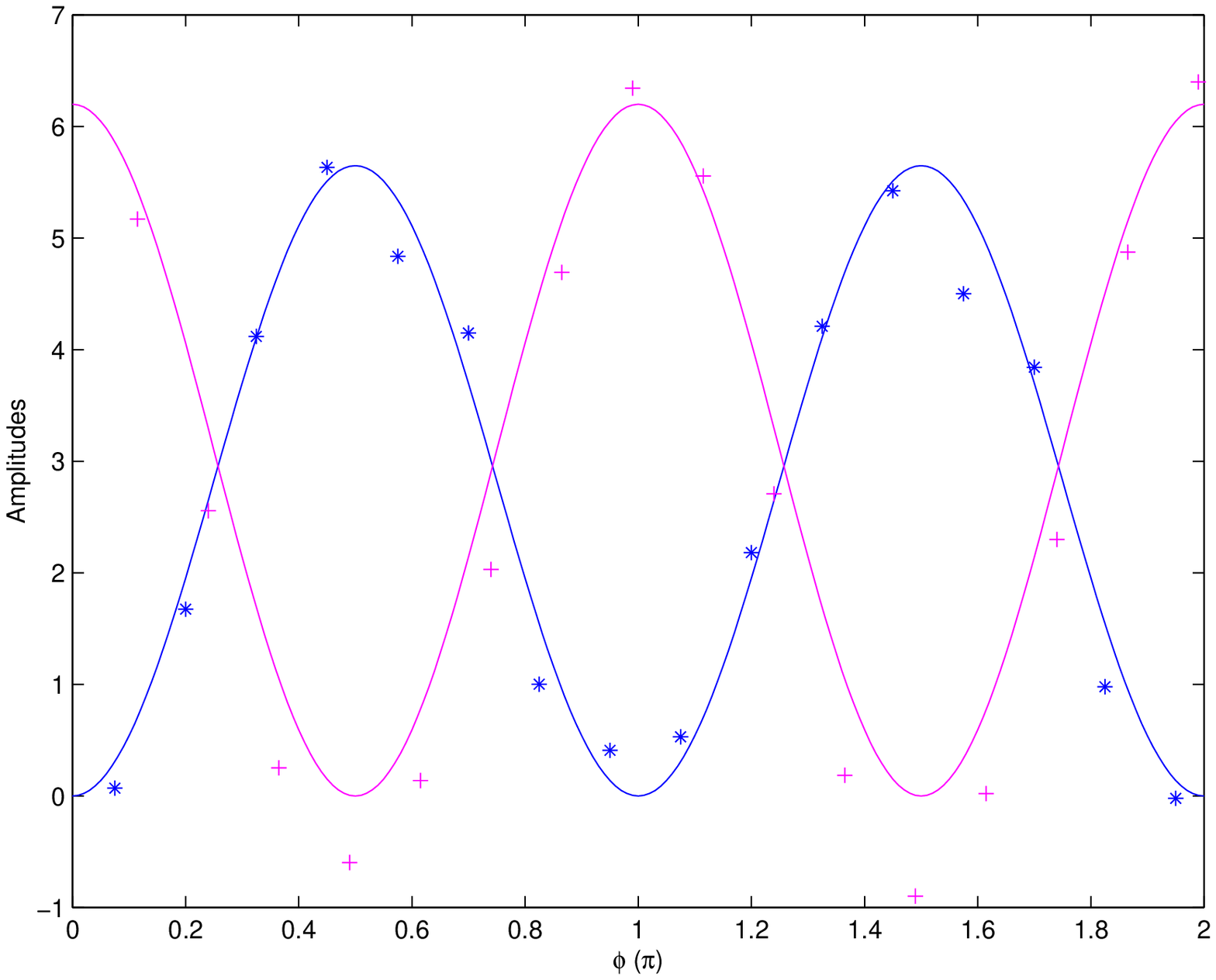}
\caption{The graph of the amplitudes of C1 and C3 vs. $\varphi=
Jt/\sqrt{2}$. The amplitudes have arbitrary units. The data for C1
are marked by "+", and are fitted as $A_{1}\cos^{2}\varphi$; the
data for C3 are marked by " * ", and are fitted as
$A_{3}\sin^{2}\varphi$, where $A_{1}=6.20$ and $A_{3}=5.65$.}
\label{pst_cb}
\end{figure}

\begin{figure}
\includegraphics[width=5.5in]{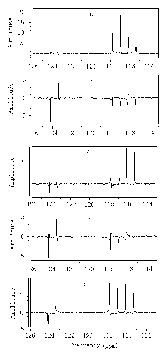}
\caption{The NMR spectra for the implementation of the perfect
state transfer when the initial state is $\sigma_{y}^{1}$. When
$\varphi=0$, $\varphi=\pi/2$, $\varphi=\pi$, $\varphi=3\pi/2$, and
$\varphi=2\pi$, the system lies in $\sigma_{y}^{1}$ (the initial
state), $-
 \sigma_{z}^{1}\sigma_{z}^{2}\sigma_{y}^{3}$, $\sigma_{y}^{1}$, $-
 \sigma_{z}^{1}\sigma_{z}^{2}\sigma_{y}^{3}$, and $\sigma_{y}^{1}$,
 shown as Figs. (a-e), respectively. Fig. (a) is the reference spectrum
used to calibrate the phases of the signals in Figs.
(b-e).}\label{pstcb1}
\end{figure}

\begin{figure}
\includegraphics[width=5.5in]{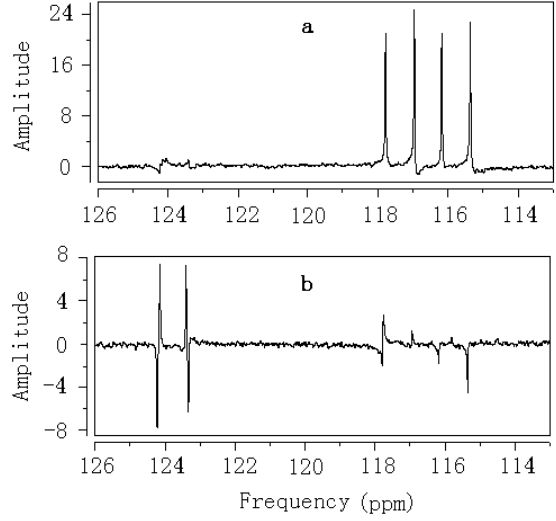}
\caption{The NMR spectra for the implementation of the perfect
state transfer when the initial state is $\sigma_{x}^{1}$. When
$\varphi=0$ and  $\varphi=\pi/2$, the system lies in
$\sigma_{x}^{1}$ (the initial state) and $-
 \sigma_{z}^{1}\sigma_{z}^{2}\sigma_{x}^{3}$
 shown as Figs. (a-b), respectively. Fig. (a) is the reference
 spectrum. There is a $\pi/2$ phase difference between the signals
 in Fig. (a) and Fig. \ref{pstcb1}(a).}\label{pstca}
\end{figure}

  We represent the results of the implementation by NMR spectra. When $\varphi$
changes, the amplitudes of C1 and C3 change as $\cos^{2}\varphi$
and $\sin^{2}\varphi$, respectively. When the initial state is
chosen as $\rho_{ini A}$, the experimental results are shown as
Fig. \ref{pst_cb}. The data for C1 are marked by "+", and are
fitted as $A_{1}\cos^{2}\varphi$; the data for C3 are marked by "
* ", and are fitted as $A_{3}\sin^{2}\varphi$. The two constants
$A_{1}=6.20$ and $A_{3}=5.65$, with arbitrary units. The
experimental results, barring two data for C1, show a good
agreement with the theoretical expectations. Figs. \ref{pstcb1}
show the spectra when the state transfers occur. When $\varphi=0$,
$\varphi=\pi/2$, $\varphi=\pi$, $\varphi=3\pi/2$, and
$\varphi=2\pi$, the system lies in $\sigma_{y}^{1}$ (the initial
state), $-
 \sigma_{z}^{1}\sigma_{z}^{2}\sigma_{y}^{3}$, $\sigma_{y}^{1}$, $-
 \sigma_{z}^{1}\sigma_{z}^{2}\sigma_{y}^{3}$, and $\sigma_{y}^{1}$,
 shown as Figs. \ref{pstcb1}(a-e), respectively. The experimental
 results, barring the signals of C1 in Figs. \ref{pstcb1}(b) and (d) of which amplitudes are
shown in Fig.\ref{pst_cb}, agree with the theoretical expectation
quite well. Theoretically, the signals of C1 in Figs.
\ref{pstcb1}(b) and (d) should not appear. The time duration for
implementing $U_{A}$ is about 200ms, which is in the same order
with the decoherence time. Hence the decoherence time limit
results in main errors. Moreover, the imperfection of the pulses
and the inhomogeneity in the magnetic field also cause errors. The
similar results can be obtained when the initial state is chosen
as $\rho_{ini B}$. Figs. \ref{pstca} show the implementation of
the perfect state transfer when the initial state is $\rho_{ini
B}$. When $\varphi=0$ and $\varphi=\pi/2$, the system lies in
$\sigma_{x}^{1}$ (the initial state) and
$-\sigma_{z}^{1}\sigma_{z}^{2}\sigma_{x}^{3}$,
 shown as Figs. \ref{pstca}(a-b), respectively.

\section{Conclusion}
   We have simulated the three- spin XY chain using liquid NMR.
 Through defining proper operators, we use the theories
of angular momentum to decompose the evolution caused by XY-
coupling into a series of factors that can be realized by rf
pulses and $J$- couplings. Such an analogue can be helpful for
solving the general problems on the Heisenberg chain.
As an example for the application of the XY- chain in quantum
computation, the perfect state transfer algorithm is realized in
the chain.

  The evolution caused by XY- couplings can be represented by single- spin operations
and the $J$- couplings, although there are no real XY- couplings
in liquid NMR. In the sample used in our experiments, the coupling
constants are not equal to each other. However we simulate the
equal couplings in the XY- chain through choosing the proper
evolution time. For the PST in more than three spin networks, the
coupling strengths are needed to be designed in a proper manner
\cite{Christandl}. Our work has shown that such couplings are easy
to simulate in NMR. All these facts represent the powerful
function of the liquid NMR in implementing quantum computation.

\section{Acknowledgment}
 This work is supported by the National Natural Science
Foundation of China under Grant No. 10374010, 60073009, 10325521,
the National Fundamental Research Program Grant No. 001CB309308,
the Hang-Tian Science Fund, the SRFDP program of Education
Ministry of China, and  China Postdoctoral Science Foundation.
J.-F. Zhang is also grateful to Dr. Peng Zhang of the Institute of
Theoretical Physics in the Chinese Academy of Science and Prof.
Jiangfeng Du of the University of Science and Technology of China
for their helpful discussions.

\newpage
\newpage
\end{document}